\documentclass[12pt]{iopart}
 \usepackage{iopams}  
 \newcommand{\pd}[2]{\frac{\partial #1}{\partial #2}}

\newcommand{\sint}{\int_{-\infty}^\infty ds\,}
 \newcommand{\be}{\begin{equation}}
  \newcommand{\ee}{\end{equation}}

\begin{document}
\rightline{NSF-KITP-09-125}

 \title[Boundary Integrals and MRE for Graphs]
 {The Boundary-Integral Formulation and 
Multiple-Reflection Expansion  for the Vacuum Energy of Quantum 
Graphs}
  
 \author[S. A. Fulling]{S. A. Fulling}
 \address{Departments of Mathematics and Physics,
  Texas A\&M University, College Station, TX, 77843-3368 USA}

 \ead{fulling@math.tamu.edu}

 \begin{abstract}
 Vacuum energy and other spectral functions of Laplace-type 
differential operators have been studied approximately by 
classical-path constructions and more fundamentally by boundary 
integral equations.
 As the first step in a program of elucidating the connections 
between these approaches and improving the resulting calculations,
 I show here how the known solutions for Kirchhoff quantum graphs 
emerge in a boundary-integral formulation.
  \end{abstract}
 

 \maketitle
 

\section{Introduction} \label{sec:intro}

 \subsection{Vacuum energy density and cylinder kernels}
  \label{ssec:vacen}
 The energy density of a quantum field in its vacuum state
 (or a state of fixed temperature), given 
particular time-independent external conditions, is of interest in 
cosmology, hadronic physics, and soft condensed matter physics 
(where the fluctuations involved are of  thermal rather than 
quantum origin). Most famously, electromagnetic vacuum energy gives 
rise to the observed Casimir force between conducting bodies (and 
its counterpart for dielectrics, the Lifshitz force).\footnote
 {In lieu of a long list of references on vacuum energy, 
 I recommend  
 the recent special issues of \emph{New Journal of Physics} 
\cite{NJP} and \emph{Journal of Physics A} \cite{qfext08}
 and the bibliographies of the papers therein.}
The study of gravitational effects (or their negligibility, even 
in the presence of divergences)
requires  detailed knowledge of all components of the 
 stress-energy-momentum tensor of the field as functions of 
position in space.  This local information is also helpful in 
addressing such 
 subjects of continued investigation and debate as
 the occasionally counterintuitive signs of the calculated forces, 
and the meaning of the divergences resulting from  idealized 
boundary conditions.

 A convenient way to study vacuum energy is to insert (usually 
temporarily) an exponential ultraviolet cutoff in the spectral 
expansions of the stress tensor or the total energy.
 For the fiducial toy model, a scalar field, this procedure
 leads quickly to the study of the \emph{cylinder} (or Poisson) 
\emph{kernels},
 which  are Green functions for certain elliptic boundary value 
problems in one higher spatial dimension  \cite{lukosz2,BH,fuliowa}.
In cavities or billiards the cylinder kernel lends itself to the 
same kind of semiclassical or optical expansion that is more often 
 (e.g., \cite{SJ1})
applied to the Green functions for the heat, Schr\"odinger, or 
resolvent kernels.
 (I hope to publish elsewhere a detailed justification and 
development of this assertion.)
 For rectangular cavities this construction reduces to the classic 
``method of images'' and yields exact results \cite{rect,liu}, 
revealing, for example, that the divergent pressure on one face of 
a rectangle is related to the divergent energy density parallel to 
the adjacent faces.  This insight is clearly pertinent to the vexed 
question of the vacuum pressure on a conducting sphere, where 
perpendicular boundary surfaces do not exist.
 Unfortunately, in the presence of curved surfaces 
 (or, \emph{a fortiori}, edges and corners)
 the optical construction is no longer exact. Preliminary 
calculations show that the leading semiclassical approximation is 
not adequate to understand the radial pressure in a sphere, and 
that in problems with curved boundaries the construction of 
 higher-order approximations is blocked by problems of the same 
type that arose in connection with the heat kernel 
\cite{McAO1,McA1}.
 When the boundary is not smooth,  the leading optical 
approximation is completely unacceptable until supplemented by 
\emph{diffractive} contributions (e.g., \cite{keller2,BPS}), which are 
hard to calculate in generality.

 \subsection{Boundary integrals and multiple reflection}
  \label{ssec:bimr}
 Although higher-order optical approximations for smooth curved 
boundaries and diffractive approximations for nonsmooth ones can 
sometimes be found by trial and error and verified \emph{a 
posteriori},
 a more systematic approach is desirable.
 In principle, such an approach is available in the standard 
mathematical literature on partial differential and integral 
equations \cite{lovitt,smithies,tricomi,folland,rubinstein}.
 A partial differential equation with boundary conditions can be 
reduced to an integral equation on the boundary.
 Up to some convergence issues touched upon lightly in  
\ref{app:converg}, the solution of the integral equation by the 
methods of Volterra or Fredholm is constructive.
The result is a \emph{multiple reflection expansion} (MRE) expressing 
 the Green function of the original problem as a series 
 $ \sum_{N=0}^\infty G_N\,$,
 where $G_N$ is an $N$-fold integral over the boundary;
 the solution is formally a ``sum'' over all paths from the source 
point to the field point, each path bouncing off the boundary $N$ 
times (in general nonspecularly).
 This construction was famously applied to physical problems by 
Balian and Bloch \cite{BB1,BB2,BB3}.  (See also \cite{PNB} in the 
case of the heat equation.)
 The more familiar (and simpler) semiclassical or optical 
approximations
 (involving only specular reflections)
  for the heat, Schr\"odinger, and resolvent equations 
 emerge from the MRE when the integrals are approximated by 
steepest descent or stationary phase.
 In the case of the cylinder kernel treated directly by MRE, 
 the appropriate approximation method is not so obvious;
 cylinder kernels can also be obtained from the other kernels by 
integral transformations, but at the cost of integrating parameters 
over values where the validity of the (e.g.)\ stationary-phase 
approximation is dubious.
 This circumstance calls for a careful examination of the 
implications of the MRE  for the representation and 
approximation of cylinder kernels for curved or nonsmooth 
boundaries --- and, indeed, at the first step, even for flat smooth 
boundaries.

 \subsection{Quantum graphs}
  \label{ssec:qgraphs}
 The present paper is a preliminary foray with primarily 
pedagogical intent.  The application of the MRE to vacuum energy 
 needs to be  
studied first in the simplest case, one spatial dimension.
 But a one-dimensional cavity is just an interval, too trivial to 
occupy one's attention for long.
 A generalization, which combines the ease of exact solution of a 
one-dimensional problem with some of the nontrivial properties of 
multidimensional systems, is the concept of a quantum graph.

 A \emph{quantum graph} \cite{kuchment1,GS}  
 is a Riemannian one-complex --- that is, a network of edges and 
vertices equipped with a natural notion of arc length on each edge.
 There follows a natural definition of the Laplacian operator as 
$d^2/dx^2$ on each edge, supplemented by boundary conditions at 
each vertex to make the operator self-adjoint.
 In particular, the most natural and popular boundary conditions 
are the \emph{Kirchhoff conditions}:
 Functions in the domain of the operator are required to be 
continuous at the vertices, while the outward derivatives of a 
function on all the edges incident on a given vertex are required 
to sum to zero.
 (A vertex attached to only one edge is thus a Neumann endpoint in 
the usual sense.)

 On a Kirchhoff quantum graph the semiclassical approximation again 
reduces to the (exact) method of images, which is, nevertheless, 
quite intricate to execute.  Solutions for the heat and resolvent 
equations have been known for some time \cite{roth,KSm,BG,KScsnobd}.
 The corresponding construction for the cylinder kernel has been 
extensively investigated by Wilson et al.\
\cite{wilson,FKaW,FKuW,FW,BHW}.
 Here I show how the same results can be obtained from an MRE, 
temporarily ignoring the fact that the simpler ansatz is already 
exact.  The connection is not exactly trivial, so the exercise will 
be useful in tackling more serious models later.
 (It also enables us to get out of the way some complications 
peculiar to the one-dimensional case, associated with the logarithmic
 form of the integral kernel~\eref{o1}.)

 \Sref{sec:notat} sets up some machinery.
 The next four sections consider four increasingly complicated 
scenarios: an interval with one endpoint, a graph with one vertex 
but arbitrarily many edges, an interval with two endpoints, and 
finally a general quantum graph (with finitely many edges, finite
in length).
In each case the MRE calculations are carried only to the point 
where it is obvious how to continue them and match them up with 
known results. 
 The main point becomes visible in the third case 
(\sref{sec:case3}), where a nontrivial boundary integral equation 
must be derived and iteratively solved for the first time.
 Some necessary integrals are evaluated in \ref{app:integrals}.
 Of greater interest is \ref{app:converg}, where the mathematical 
status of the multiple-reflection series is discussed, and the 
possible utility of the exact Fredholm solution in the situation of 
finite temperature is pointed out.
The machinery for finite temperature is set up in \ref{app:temper}.
 
\section{Notation}\label{sec:notat}

Let $\Gamma$ be a quantum graph, and let
  $\tilde\Gamma = \mathbf{R}\times\Gamma$ be the 
corresponding ``Euclideanized space-time''\negthinspace.  
 Variables $x$, $y$, $\ldots$ stand for points 
in $\Gamma$ and variables $s$, $t$, $\ldots$ for real numbers.
I consider here only the standard Kirchhoff boundary conditions.

The Green function for the Poisson equation in $\mathbf{R}^2$ is
\begin{eqnarray}
  G_0(t,x;s,y) &= -\,\frac1{2\pi} \ln \left (\frac r{r_0}\right ) 
 \nonumber \\
&= -\,\frac1{4\pi} \ln[(t-s)^2 +(x-y)^2] +C. 
\label{o1} \end{eqnarray}
Let $G(t,x;s,y)$ be the Green function for the Poisson equation in 
$\tilde\Gamma$.
Then the cylinder kernels in $\tilde\Gamma$ are
\begin{eqnarray} \overline T(t,x,y) = -2 G(t,x;0,y), \label{o2} \\
T(t,x,y) = -2 \,\pd Gt(t,x;0,y). \label{o3}
\end{eqnarray}
$G_0$ is defined only modulo the indicated scale ambiguity
 (the arbitrary constants $r_0$ and $C$), 
but $T_0$ and the spatial derivatives of $G_0$ are unique.
Henceforth we take $C=0$.

 For later use note
 \be
 \pd{G_0}x(t,x;s,y) = -\,\frac1{2\pi}\,
 {x-y\over (t-s)^2+ (x-y)^2} = -\,\pd{G_0}y\,.
 \label{deriv}\ee
 Related to this function is the well known 
 distributional identity
 \be
 \lim_{x\downarrow0} \frac1{\pi}\, \frac x{(t-s)^2 +x^2} =
 \delta(t-s), \label{delta}
 \ee
 which will be used repeatedly in what follows.

\section{Case 1: $\Gamma=\mathbf{R}^+$} \label{sec:case1}

We will solve the Poisson equation on the half-plane with Neumann 
boundary at $y=0$.

Make the ansatz
\begin{eqnarray} G = G_0+\gamma, \label{o4} \\
\gamma(t,x;s_0,y_0) = \sint G_0(t,x;s,0)\mu_0(s), \label{o5}
\end{eqnarray}
where $\mu_0$ has a dependence on $(s_0,y_0)$ that will be notationally 
suppressed.
The integral in \eref{o5} is over the boundary $\{(s,y)\colon y=0\}$, 
 and the subscript on $\mu$ refers to that value of~$y$.

We calculate from \eref{deriv}
\begin{eqnarray} \pd \gamma x(t,0;s_0,y_0) 
&= -\sint \frac{x-y}{2\pi}\left.[(t-s)^2+(x-y)^2]^{-1}\right|_{x=y=0} 
\mu_0(s) \nonumber  \\
&=0. \label{o6}\end{eqnarray}
As always in the boundary-integral method, we must make a careful 
distinction between the value of such an integral exactly on the 
boundary ($x=0$) and the limit of the integral as the boundary is 
approached from the interior.  For the latter we have by 
\eref{deriv} and \eref{delta}
\begin{eqnarray}  \pd \gamma x(t,x;s_0,y_0)
&= -\sint \frac x{2\pi}[(t-s)^2+x^2]^{-1}\mu_0(s) \nonumber \\
&\to -\,\frac12 \mu_0(t)\quad\mbox{as $x\downarrow 0$,}
\label{o7}\end{eqnarray}
and it is this object that must be chosen to satisfy the Neumann 
boundary condition:
\[ 0= \pd Gx (t,0^+;s_0,y_0) = \pd {G_0}x + \pd \gamma x \]
implies
\be
 \mu_0(t) = 2\,\pd {G_0}x (t,0^+;s_0,y_0)
= \frac{y_0}{\pi} [(t-s_0)^2 + y_0{}\!^2]^{-1} \label{o8}\ee
and hence
\be \gamma(t,x;s_0,y_0) = -\, \frac{y_0}{4\pi^2}
\sint \ln[(t-s)^2+x^2][(s-s_0)^2 + y_0{}\!^2]^{-1}.
\label{o9}\ee

But from the method of images we know a more elementary formula for 
$\gamma$:
\be
 \gamma(t,x;s_0,y_0) =
G_0(t,x;s_0,-y_0) = -\,\frac1{4\pi} \ln[(t-s_0)^2+(x+y_0)^2].
\label{o10}\ee
The equivalence of \eref{o9} and \eref{o10} is shown in  
\ref{app:integrals}.

\section{Case 2: Infinite star graphs}
\label{sec:case2}

Let $d_v$ be the number of edges meeting at the central vertex.
On edge $j$ there is a coordinate $x$, sometimes written $x_j\,$,
equal to $0$ at the vertex.  Without loss of generality we can take 
$y_0$ to be located on edge $j=1$. 

In analogy with \eref{o4} and \eref{o5}
  we construct the Green function in the form
\be  G_j = \cases{ G_0 +\gamma_j&if $j=1$, \cr
                    \gamma_j&if $j\ne1$, \cr} 
 \label{o11}\ee
 \begin{eqnarray}
\fl\gamma_j(t,x;s_0,y_0) &= \sint G_0(t,x;s,0) \mu_j(s) + \sint 
\pd{G_0}y(t,x;s,0)\nu_j(s) \nonumber  \\
&= -\,\frac1{4\pi}\sint \ln[(t-s)^2+x^2]\mu_j(s) \nonumber\\ 
&\qquad{}+ \frac x{2\pi} \sint [(t-s)^2+x^2]^{-1}\nu_j(s). 
\label {o12} \end{eqnarray}
(Here the dependence of $\mu$ and $\nu$ on the vertex $v$ is suppressed, 
along with that on $(s_0,y_0)$.)

The Kirchhoff boundary conditions are
 \begin{eqnarray}
 G_j(t,0;s_0,y_0) = G_k(t,0; s_0,y_0) \quad\mbox{for all $j,k$}, 
 \label{o13} \\
\sum_{j=1}^{d_v} \pd {G_j}x(t,0; s_0,y_0) =0. 
 \label{o14}\end{eqnarray}
They translate into
 \begin{eqnarray}
 \gamma_j(t,0^+;s_0,y_0) = \gamma_k(t,0^+; s_0,y_0) 
 \quad\mbox{for all $j,k\ne1$}, 
 \label{o15} \\
\gamma_1(t,0^+;s_0,y_0) = \gamma_k(t,0^+; s_0,y_0) - G_0(t,0;s_0,y_0)
\quad\mbox{(for any $k\ne1$)},
  \label{o16} \\
\sum_{j=1}^{d_v} \pd {\gamma_j}x(t,0^+; s_0,y_0) =
-\,\pd {G_0}x(t,0;s_0,y_0).
  \label{o17} \end{eqnarray}
The symmetries of these equations suggest the further ans\"atze
 \begin{eqnarray}
 \mbox{All $\mu_j$ are equal}, \label{o18} \\
\mbox{All $\nu_j$ are equal except $\nu_1\,$}, \label{o19} \\
 \sum_{j=1}^{d_v} \nu_j = 0 . \label{o20} 
\end{eqnarray}
Then \eref{o15} is satisfied, and \eref{o16} becomes,
 by  analogy with \eref{o7},
\be
 \nu_1(t) -\nu_k(t)  =
-2G_0(t,0;s_0,y_0) = \frac1{2\pi} \ln[(t-s_0)^2+y_0{}\!^2]. 
\label{o21}\ee  
Sum \eref{o21} over all $k\ne1$ and use \eref{o20}:
\[\frac{d_v-1}{2\pi} \ln[(t-s_0)^2+y_0{}\!^2] =
(d_v-1)\nu_1 - \sum_{k\ne1} \nu_k = d_v\nu_1 - \sum_{j=1}^{d_v} \nu_k 
= d_v \nu_1\,.\]
Thus
\be\nu_1(t) = \left (1-\frac1{d_v}\right ) 
 \frac1{2\pi} \ln[(t-s_0)^2+y_0{}\!^2],
\label{o22}\ee
and then
\be\nu_k(t) = -\,\frac1{2\pi d_v} \ln[(t-s_0)^2+y_0{}\!^2].
\label{o23}\ee
Finally, to impose \eref{o17} and find $\mu_j$ we note from 
 \eref{o12} that
\[\fl
 \pd{\gamma_j}x (t,x;s_0,y_0) =
-\,\frac x{2\pi}\sint [(t-s)^2+x^2]^{-1} \mu_j(s) +
\sint \pd{^2G_0}{x\,\partial y} (t,x;s,0) \nu_j(s),\]
and hence from \eref{o17} and \eref{o20} 
\[-\,\pd{G_0}x (t,0;s_0,y_0) = -\lim_{x\downarrow0}
\frac x{2\pi}\sint [(t-s)^2+x^2]^{-1} \sum_{j=1}^{d_v} \mu_j(s).\]
It follows by \eref{o18},  \eref{o1}, and \eref{delta} that
\be\mu_j(t) = \frac{y_0}{\pi d_v} [(t-s_0)^2+ y_0{}\!^2]^{-1}.
\label{o24}\ee

Inserting \eref{o22}--\eref{o24} into \eref{o12} we arrive at
\begin{eqnarray} \gamma_j(t,x;s_0,y_0)
  &= -\frac{1}{4\pi^2d_v}\sint  y_0\,
\frac{\ln[(t-s)^2+x^2]}{(s-s_0)^2+y_0{}\!^2}  \nonumber \\
&\qquad{}+\left (\delta_{j1}-\frac1{d_v}\right ) \frac 1{4\pi^2} 
\sint x\,\frac{\ln[(s-s_0)^2+y_0{}\!^2]}{(t-s)^2+x^2} \, .
\label{o25} \end{eqnarray}
The two integrals in \eref{o25} are identical except for the interchange
$(t,x)\leftrightarrow (s_0,y_0)$.
Moreover, at the end of the previous section, and in \eref{A3},
 we observed that
\be\frac{y_0}{\pi} \sint \frac{\ln[(t-s)^2+x^2]}{(s-s_0)^2 +y_0{}\!^2}
= \ln [(t-s_0)^2 + (x+y_0)^2]; 
 \label{o26}\ee
this function is actually invariant under that interchange.
(In \eref{o26} it is assumed that $x$ and $y_0$ are positive.)
So we finally arrive at
\be
 \gamma_j(t,x;s_0,y_0) =\left (\delta_{jl} - \frac2{d_v}\right ) \frac1{4\pi}
\ln[(t-s_0)^2 + (x+y_0)^2],  
 \label{o27}\ee
where we now allow for $y_0$ to be located on any edge, $l$.
This is equivalent to known results for quantum star graphs
 (e.g., \cite[Sec.~3B]{KSm}, \cite[Ch.~3]{wilson}).
From \eref{o3}, the formula for the cylinder kernel $T$ is
\be
 T_j^l(t,x,y) = \delta_j^l \, \frac{t/\pi}{t^2+(x-y)^2}
+\left (\frac2{d_v} - \delta_j^l\right ) \frac{t/\pi}{t^2+(x+y)^2}\,, 
 \label{o28}\ee
which is equation (35) of \cite{snobd} with the Robin 
parameter $\alpha=0$.

\section{Case 3: $\Gamma=(0,L)$} \label{sec:case3}

I write $\partial/\partial n$
 for an \emph{inward} normal derivative in the usual 
sense of bounded domains.  In the graph context such a derivative is 
\emph{outward} from a vertex.

In the present model the boundary has two parts, 
$\{(s,y)\colon y=0\}$ and $\{(s,y)\colon y=L\}$, 
the boundary condition is still Neumann,
and the obvious ansatz 
is \eref{o4} with
\be\fl\qquad
 \gamma(t,x;s_0,y_0) = \sint G_0(t,x;s,0)\mu_0(s) +
\sint G_0(t,x;s,L)\mu_L(s). 
 \label{o29}\ee
Evaluate the derivative:
\begin{eqnarray}
  \pd{\gamma}x(t,x;s_0,y_0)  
&=-\frac1{2\pi} \sint \frac x{(t-s)^2+x^2}\, \mu_0(s) 
 \nonumber \\
&\qquad{}-\frac1{2\pi} \sint \frac {x-L}{(t-s)^2+(x-L)^2}\, \mu_L(s).
\label{o30}\end{eqnarray}
The limits of the normal derivatives from inside are
 \begin{eqnarray}
\fl\quad \pd{\gamma}n(0^+) &\equiv \pd{\gamma}x(t,0^+;s_0,y_0) 
= -\,\frac12 \,\mu_0(t) +\frac1{2\pi} \sint \frac L{(t-s)^2+L^2} 
\,\mu_L(s), \label{o31} \\
\fl\quad \pd{\gamma}n(L^-) &\equiv -\, \pd{\gamma}x(t,L^-;s_0,y_0) 
= -\,\frac12 \,\mu_L(t) +\frac1{2\pi} \sint \frac L{(t-s)^2+L^2} 
\,\mu_0(s).
  \label{o32} \end{eqnarray}
(The sign constraint in \eref{o7} becomes $L-x\downarrow0$ 
 in going from \eref{o30} to~\eref{o32}.)
The appropriate boundary conditions are
\be
 \pd{\gamma}n(0^+) = -\, \pd{G_0}x(t,0;s_0,y_0), \quad
 \pd{\gamma}n(L^-) = +\, \pd{G_0}x(t,L;s_0,y_0) . 
 \label{o33}\ee
From these relations we obtain the basic boundary-integral equations of 
the problem,
 \begin{eqnarray}
 \mu_0(t) &= \frac1{\pi} \,\frac{y_0}{(t-s_0)^2 +y_0{}\!^2}
+ \frac1{\pi}\sint \frac L{(t-s)^2+L^2}\, \mu_L(s), 
 \nonumber \\
\mu_L(t) &= \frac1{\pi} \,\frac{L-y_0}{(t-s_0)^2 +(L-y_0)^2}
+ \frac1{\pi}\sint \frac L{(t-s)^2+L^2}\, \mu_0(s). 
 \label{o34} \end{eqnarray}

More abstractly, the system \eref{o34} is
\be
 \mu_v(t) = \frac1{\pi} \, \frac{d(v,y_0)}{(t-s_0)^2 + d(v,y_0)^2}
+ \frac1{\pi} \sint \frac L{(t-s)^2+L^2} \, \mu_{\overline v}(s),
\label{o35}\ee
where $d(v,y_0)$ is the distance from vertex $v$ to $y_0\,$,
and $\overline v$ is the \emph{other} vertex.
Still more abstractly, fix a vertex $v_*\,$; then
\be
 \mu_{v_*}(t) = 2 \, \pd{G_0}n(t,v_*;s_0,y_0) +
2\int_{\partial\tilde\Gamma} \pd{G_0}n(t,v_*;s,v)\mu_v(s), 
 \label{o36}\ee
where a sum over $v$ is implicit in the integration over the total 
boundary $\partial\tilde\Gamma=\mathbf{R}\times\partial\Gamma$ of the graph,
 and it happens, as in \eref{o6}, that
\be \pd{G_0}n(t,v_*;s,v_*) =0 . \label{o37}\ee
Even more abstractly, the integral equation has the form
\be \mu = g_0 + K\mu, \label{o38}\ee
so that formally 
\be \mu = (1-K)^{-1}g_0 \sim (1 + K + K^2 + \cdots) g_0\,.
\label{o39}\ee

Postponing to \ref{app:converg}
  the issue of the convergence of the Neumann 
series \eref{o39}, we examine its zeroth term, 
 which produces the first-order 
(not leading) term in the series for $G$.
That is, one drops the integrals in \eref{o34} and 
 substitutes \eref{o34} into \eref{o4} and \eref{o29}:
 \begin{eqnarray}\fl
 G(t,x;s_0,y_0)&{} - G_0(t,x;s_0,y_0) \nonumber \\ 
&= -\,\frac1{4\pi^2} \left \{ \sint 
\ln[(t-s)^2+x^2]\frac{y_0}{(t-s_0)^2+y_0{}\!^2} \right . \nonumber\\
&\qquad{} + \left .\sint 
\ln[(t-s)^2+(L-x)^2]\frac{L-y_0}{(t-s_0)^2+(L-y_0)^2}\right \} 
 \nonumber \\
&=- \,\frac1{4\pi}\left \{\ln[(t-s_0)^2+(x+y_0)^2] 
+ \ln[(t-s_0)^2 + (2L-x-y_0)^2] \right \},
\label{o40} \end{eqnarray}
where \eref{o26} has been used (with $L-x>0$, $L-y_0>0$).
Clearly, \eref{o40} comprises the single-reflection terms in the standard 
solution by the method of images
(images at $-y_0$ and $2L-y_0$).

The first-order term in \eref{o39}
  should therefore yield the two-reflection terms in the image solution. 
  Substituting the zeroth-order $\mu$
(the first terms in \eref{o34}) into the integrals in \eref{o34}, one gets
  \begin{eqnarray}\fl  \null\hfill                   
  \mu_0(t) - \frac1{\pi} \,\frac{y_0}{(t-s_0)^2 +y_0{}\!^2}
\hfilneg &= \frac1{\pi^2}\sint \frac 
{L(L-y_0)}{[(t-s)^2+L^2][(s-s_0)^2+(L-y_0)^2]} \nonumber\\
&= \frac1{\pi} \, \frac{2L-y_0 }{ (t-s_0)^2+(2L-y_0)^2}\,, 
 \nonumber \\
\fl \mu_L(t) - \frac1{\pi} \,\frac{L-y_0}{(t-s_0)^2 +(L-y_0)^2} &=
 \frac1{\pi^2}\sint \frac {Ly_0}{[(t-s)^2+L^2][(s-s_0)^2+y_0{}\!^2]}
\nonumber\\ 
 &=\frac1{\pi} \,\frac{L+y_0}{(t-s_0)^2+(L+y_0)^2} \, . 
 \label{o41}\end{eqnarray}
(For the evaluation of these integrals see \eref{A1}.)
The resulting additional terms in $G$ are 
\begin{eqnarray}\fl
 -\,\frac1{4\pi^2} \left \{\sint \ln[(t-s)^2+x^2]
\frac{2L-y_0}{(s-s_0)^2 + (2L-y_0)^2}\right . \nonumber\\
\quad{}+\left .\sint
\ln[(t-s)^2+(L-x)^2] \frac{L+y_0}{(s-s_0)^2 +(L+y_0)^2} \right \}
\label{o42} \\
= -\,\frac1{4\pi} \left \{
\ln[(t-s_0)^2+(x+2L-y_0)^2] + \ln[(t-s_0)^2 + (2L-x+y_0)^2] \right \}.
\nonumber\end{eqnarray}
Exactly as expected, these terms describe images at $y_0-2L$ and 
$y_0+2L$.

\section{Case 4: General compact Kirchhoff quantum graphs}
\label{sec:case4}

\emph{Notation:}  
 $v$ is a vertex of degree $d_v\,$;  $e$ is an edge.
Whenever considering a fixed vertex~$v$, we can assume that the edges 
are parametrized so that $x\equiv x_e= 0$ at $v$;
each incident edge ($e\in E_v$) will then have a terminal vertex $t_e$ 
at which $x_e=L_e\,$.
Let $e_0$ be the edge containing the source point $y_0\,$.

\emph{Ansatz} (generalizing \eref{o12} and \eref{o29}):
\be G_e = \delta_{ee_0} G_0 + \gamma_e\,.  
 \label{o43}\ee
For every $v$ and every   $e\in E_v$ there will be charge and dipole 
densities $\mu_{ev}(t)$ and $\nu_{ev}(t)$
(with hidden dependence on $(s_0,y_0)$ as usual).
For temporary notational purposes, given $e$ choose one of its two 
vertices, $v$, as its initial vertex.
Then 
\begin{eqnarray}\fl
\gamma_e(t,x;s_0,y_0) = \sint G_0(t,x;s,0) \mu_{ev}(s)
+\sint G_0(t,x;s,L_e) \mu_{et}(s) \nonumber \\
{}+\sint \pd {G_0}y(t,x;s,0)\nu_{ev}(s) -
\sint \pd {G_0}y(t,x;s,L_e)\nu_{et}(s).
\label{o44}\end{eqnarray}

\emph{Boundary conditions:}
Also for temporary purposes, consider a fixed $v$ 
and assume the edges 
in $E_v$ are numbered $1$, $\ldots$, $d_v\,$.
(Of course, this list may or may not include $e_0\,$.)
Then the boundary conditions at $v$ are, for all $e$ and $f$ in $E_v\,$,
 \begin{eqnarray}\fl\qquad
 \gamma_e(t,0;s_0,y_0)+\delta_{ee_0}G_0(t,0;s_0,y_0) =
\gamma_f(t,0;s_0,y_0)+\delta_{fe_0}G_0(t,0;s_0,y_0), 
 \label{o45} \\  \fl\qquad
\sum_{e=1}^{d_v} \pd {\gamma_e}x (t,0;s_0,y_0) =
-\,\pd{G_0}x(t,0;s_0,y_0) \sum_{e=1}^{d_v} \delta_{ee_0}\,.
\label{o46}\end{eqnarray}
For \eref{o46} we calculate
\begin{eqnarray}\fl \pd{\gamma_e}x(t,x;s_0,y_0) =
\sint \pd{G_0}x(t,x;s,0)\mu_{ev}(s)
 +\sint \pd{G_0}x(t,x;s,L_e)\mu_{et}(s) \nonumber\\
{}+\sint \pd{^2G_0}{x\,\partial y}(t,x;s,0)\nu_{ev}(s)
 -\sint \pd{^2G_0}{x\,\partial y}(t,x;s,L_e)\nu_{et}(s) .
\label{o47}\end{eqnarray}

\emph{Integral equation system:}
 Of the three simplifying symmetry relations
\eref{o18}--\eref{o20}, the first 
and third are fundamental (essentially characterizing the solutions for 
the Dirichlet and Neumann subspaces of the graph --- see, e.g., 
\cite{kuchment1,FKuW}).  They can immediately be generalized to the present 
situation:
 \begin{eqnarray}
 \mu_{ev} = \mu_{fv} \quad\mbox{for all $e,f\in E_v$},
  \label{o48} \\
\sum_{e=1}^{d_v} \nu_{ev} = 0. 
 \label{o49}\end{eqnarray}
However, \eref{o19} was a symmetry  that may no longer hold, because 
even if $y_0$ does not fall on any $e\in E_v\,$,
some of those edges are ``closer'' to $y_0$ than others.
As before, \eref{o48} and \eref{o49} assure that 
 all nontrivial integrals at $v$ 
in \eref{o44}--\eref{o46} cancel.
They also reduce the number of unknowns from $2d_v$ to $d_v\,$,
which is the number of independent equations in \eref{o45} and 
 \eref{o46}.
For the trivial integrals we have as usual
\be       \fl\qquad
 \pd{G_0}x(t,x;s,0) = -\, \frac{x-0}{2\pi} [(t-s)^2+(x-0)^2]^{-1}
\to -\frac12\,\delta(t-s) \quad\mbox{as $x\downarrow0$}.
\label{o50}\ee
Thus \eref{o45} reduces to
 \begin{eqnarray}\fl
  \frac12\,\nu_{ev}(t)+\sint G_0(t,0;s,L_e)\mu_{et}(s)
{}-\sint \pd{G_0}y (t,0;s,L_e)\nu_{et}(s) 
+\delta_{ee_0} G_0(t,0;s_0,y_0)  \nonumber \\
= \mbox{same with $e\to f$}. 
\label{o51} \end{eqnarray}
This set of $d_v-1$ independent equations, together with the constraint 
\eref{o49},
determines $\nu_{ev}(t)$ for all $e\in E_v\,$, if the $\nu_{et}$ and 
$\mu_{et}$ are known.
With the aid of \eref{o48}, \eref{o46} reduces to a single equation 
 to determine $\mu_{ev}(t)$:
\begin{eqnarray}\fl
  -\,\frac{d_v}2 \,\mu_{ev}(t) 
{}+\sum_{e=1}^{d_v}\sint \pd{G_0}x(t,0;s,L_e)\mu_{et}(s)
-\sum_{e=1}^{d_v}\sint \pd{^2G_0}{x\, \partial y}(t,0;s,L_e)\nu_{et}(s) 
 \nonumber\\
= -\,\pd{G_0}x(t,0;s_0,y_0) \sum_{e=1}^{d_v} \delta_{ee_0}\,.
\label{o52}\end{eqnarray}
(The last factor in \eref{o52} is equal to either $0$ or $1$,
 unless $e_0$ is a loop at $v$, in which case it equals~$2$.)

So as in \sref{sec:case3} we have in \eref{o51}--\eref{o52}
  an integral equation of the form 
\be \vec \mu = \vec g_0 + K \vec\mu , 
 \label{o53} \ee
where $\vec\mu$ is a vector-valued function whose $\sum_v d_v$ 
components are, for each $v$,
any $d_v-1$ independent choices of the quantities $\nu_{ev}$ 
(constrained by \eref{o49}) and any one 
of the (equal) quantities $\mu_{ev}\,$.
Of course, in any concrete case the ``$et$'' notation needs to be 
resolved in favor of a fixed labeling of the vertices.
In principle all integrals in the Neumann series can be formally 
 evaluated recursively, and
the results for $T(t,x;s,y)$ must be the same as those obtained 
 by Wilson  \cite{wilson,FW}.
 Convergence of the series is not immediately obvious 
 (see \ref{app:converg}).  However, convergence of the series for 
the total energy has been   proved  in \cite{BHW}.

 \ack
 This research is supported by National Science Foundation Grant
 PHY05-54849.
Much of the work was done while I
  enjoyed the hospitality and partial support of the
Institute for Mathematics and Its Applications (NSF Grant
DMS04-39734) and the Kavli Institute for
Theoretical Physics (NSF Grant PHY05-51164).
I am grateful for sympathetic audiences and occasional technical 
help to all my student research assistants 
 (especially Justin Wilson,  Zhonghai Liu, and Kevin Resil),
  the other 
members of the Texas--Oklahoma--Louisiana vacuum energy research 
consortium, and the members of the Texas A\&M
 quantum graphs research group (especially
 Brian Winn for improving the treatment of the integrals in 
\ref{app:integrals}).
 
\appendix

\section{Evaluation of integrals} \label{app:integrals}

\textbf{Lemma:}
   Let $x$ and $y_0$ be positive.  Then
\begin{eqnarray}
  \int_{-\infty}^\infty \frac{ds}{[(t-s)^2+x^2][(s-s_0)^2+y_0{}\!^2]}
&= \frac{\pi}{xy_0}\, \frac{x+y_0}{(t-s_0)^2 + (x+y_0)^2}\,,
\label{A1}\\
\int_{-\infty}^\infty \frac{(t-s)\,ds}{[(t-s)^2+x^2][(s-s_0)^2+y_0{}\!^2]}
&= \frac{\pi}{y_0}\, \frac{t-s_0}{(t-s_0)^2 + (x+y_0)^2}\,.
\label{A2} \end{eqnarray}

\textsl{Proof:}
Tediously, these integrals can be evaluated by standard methods --- 
either residues or partial fractions.
(The point $(t,x)=(s_0,y_0)$ requires special attention with a 
continuity argument.)

\textbf{Proposition:} 
 Let $x$ and $y_0$ be positive.  Then
\be
 \sint \frac{\ln[(t-s)^2+x^2]}{(s-s_0)^2 +y_0{}\!^2}
=\frac{\pi}{y_0} \,\ln[(t-s_0)^2 + (x+y_0)^2]. 
 \label{A3}\ee

\textsl{Proof:\/} 
 Note that differentiating \eref{A3} with respect to $x$ yields 
\eref{A1}, while differentiating \eref{A3} 
  with respect to $t$ yields \eref{A2}.
It follows that the integral in \eref{A3}
  equals the right-hand side of \eref{A3} 
plus a constant, $C$, that is independent of $x$ and $t$.
 The substitution $\tilde s \equiv s-s_0$ reduces the 
integral to
\be \int_{-\infty}^\infty d\tilde s \,
\frac{\ln[(t-s_0-\tilde s)^2 + x^2]}{\tilde s^ 2 + y_0{}\!^2}\,,
 \label{A4}\ee
from which it is easy to see that it is symmetric in $t$ and $s_0\,$.
Therefore, $C$ is independent of $s_0$ too.
To fix $C$ it suffices to consider the special cases
$x\to0$, $t=0$, $s_0=0$:
\begin{eqnarray}
  C&= \sint \frac{\ln s^2}{s^2+y_0{}\!^2} -\frac{\pi}{y_0}\ln 
y_0{}\!^2 \nonumber \\
&= \frac1{y_0} \int_{-\infty}^\infty d\sigma\, 
 \frac{\ln(\sigma^2y_0{}\!^2)}{\sigma^2+1}
  -\frac1{y_0} \ln y_0{}\!^2 \int_{-\infty}^\infty
  \frac{d\sigma}{\sigma^2+1} \nonumber\\
&= \frac2{y_0} \int_{-\infty}^\infty 
 \frac{\ln \sigma}{\sigma^2 +1}\, d\sigma \nonumber\\
&= \frac4{y_0}\left [\int_0^1
  \frac{\ln \sigma}{\sigma^2+1} \,d\sigma + \int_1^\infty 
\frac{\ln \sigma}{\sigma^2+1} \,d\sigma \right ] \nonumber\\
&= 0 
\label{zero} \end{eqnarray}
because the substitution $\sigma \to 1/\sigma$ converts the second term to 
the negative of the first.

\textsl{Remark:\/}  
 \eref{A3} is an instance of a formula given by Balian and 
Bloch \cite[(7.2)]{BB3},
\be
 G_0(r\tilde r') = 2 \int_S d\sigma_\alpha\, \pd{G_0(r\alpha)} {n_\alpha} 
\,G_0(\alpha r') \label{A5} \ee
where $\tilde r'$ is the image of $r'$ in the plane $S$.
They conclude it just by noting that in this case the 
boundary-integral/multiple-reflection expansion 
must agree with the image solution, assumed previously known.
One of the main motivations of the present exercise was to verify 
\eref{A5} 
directly, or, to put it differently, to \emph{derive} the image 
solution from the (more general and fundamental) boundary-integral
solution.
The symmetry of \eref{o26} in $(t,x)$ and $(s,y_0)$
 is an instance of a symmetry of 
the right-hand side of \eref{A5} noted by Hansson and Jaffe 
\cite[(A.12)]{HJ}.

\section{Convergence issues} \label{app:converg}

 \subsection{Case 3}\label{ssec:c3}
Extrapolating from \eref{o41},
  one can see that the series \eref{o39}  will not 
converge, because the terms fall off only linearly with the distance to 
the image charge.  (The corresponding series for the Dirichlet case is 
conditionally convergent because of sign alternation.)
The series for $G$, starting with \eref{o40} and \eref{o42},
  is even worse: the terms grow logarithmically.

  In studying vacuum energy we are usually interested in derivatives of 
$\overline T$.  Their series converge better, because the differentiations 
build up powers of $(t-s)^2+ (y_0+\cdots)^2$ in the denominators.
One can say that the original series converges distributionally with 
respect to  test functions possessing sufficiently many antiderivatives 
that are also test functions (in the simplest case, test functions 
orthogonal to the constant functions).

The standard test for convergence of a Neumann series like \eref{o39}
  is that 
the norm of $K$ as an operator from some Banach space into itself be 
less than~$1$.
In our case (see \eref{o34}--\eref{o36}) the kernel function is built from
\be K_0(t,s) = \frac1\pi \,\frac L{(t-s)^2 +L^2}\,,  
 \label{A6}\ee
so that
\be
 \sint K_0(t,s) = 1 = \int_{-\infty}^\infty dt\,K_0(t,s). 
 \label{A7} \ee
It follows that $K_0$ has norm $1$ as an operator in either 
 $L^1(\mathbf{R})$ or $L^\infty(\mathbf{R})$.
Furthermore, the iterated kernels have the same general form and the 
same norm; for example,
\be
 K_0{}\!^2(t,s_0) \equiv \sint K_0(t,s)K_0(s,s_0) = 
\frac1\pi\, \frac{2L}{(t-s_0)^2+(2L)^2}
  \label{A8} \ee
by \eref{A1}.
So, in the multiple-reflection expansion we are operating right on the 
circle of convergence in the plane of  a formal parameter 
multiplying~$K$
 (``marginal convergence'').
Actually, one apparently needs \eref{A8} to complete the abstract proof 
that the operator $L^1$-norm is $1$, not possibly something smaller.
The foregoing remarks about conditional and derivative convergence 
show, however, that the latter possibility does not hold here.

Note in passing that $K_0$ is certainly not trace class or 
Hilbert--Schmidt, since 
$\sint K_0{}\!^2(s,s) $ is divergent.
 (The full operator $K$ of \eref{o35} has the matrix-valued kernel
 \[ K = \left( \begin{array}{cc}
  0& K_0 \\ K_0 & 0 \end{array}\right),
 \]
 which, being totally off-diagonal, has a vanishing trace in the 
naive sense that
 \[ \sum_v \sint K_{vv}(s,s) =0, \]
 but that does not make it trace class.)
 The Fredholm solution is worse than divergent, because even the 
individual terms in the numerator and denominator series do not exist.
Of course, this is to be expected for a convolution integral 
equation on an infinite interval.
 (This issue does not arise in the standard literature 
\cite{BB1,BB2,BB3,lovitt,smithies,tricomi,folland,rubinstein}, 
 which concentrates on compact boundaries.)
 If the graph were to be studied at finite temperature
 (\ref{app:temper}), the 
interval would be finite and the Fredholm theory would be 
applicable, possibly providing better convergence than that of the 
Neumann series.

  \subsection{Case 4}\label{ssec:c4}
 The foregoing argument does not readily extend to the integral 
equation system \eref{o53}, because the undifferentiated $G_0$ 
appearing twice in \eref{o51} is not $L^1$ (nor $L^\infty$) in the 
``time'' variables. 

 At first glance it is not obvious even that the individual terms 
in the iterative solution will converge. Inspection of 
 \eref{o51}--\eref{o52}, together with 
 \eref{deriv},  \eref{A1}, and \eref{A3},
  shows by induction, however, that each term in $\nu_{ev}(t)$
 is logarithmic like $G_0$ and each term in $\mu_{ev}(t)$ has 
behavior like $\partial G_0/\partial x$ as $t\to\infty$.
(There are always enough spatial derivatives to avert a disaster.)
 It seems likely that, as in Case~3, one is sitting on the edge of 
a region of convergence, and  the convergence of $\gamma_e$
 can be improved by taking spatial derivatives.
 (See also the plausibility argument and numerical verification
  in~\cite{FKaW}.)
 However, it is not possible to prove convergence by inspection 
without control over the number of terms in each order of the 
series and the magnitudes of their numerical coefficients.

 On the question of convergence of the series, 
 some improvement is achieved by changing 
variables from $\nu$ to $\nu'$.  Equations \eref{o51} and 
\eref{o52} can be rewritten
 \begin{eqnarray}\fl
  \nu_{ev}'(t)+2\sint \pd{G_0}t(t,0;s,L_e)\mu_{et}(s)
 \nonumber \\
\null\qquad{}-2\sint \pd{^2G_0}{t\,\partial y} (t,0;s,L_e)\nu_{et}(s) 
+2 \pd{G_0}{t}(t,0;s_0,y_0)\delta_{ee_0}  \nonumber \\
= \mbox{same with $e\to f$}, 
\label{a51} \end{eqnarray}                               
 \begin{eqnarray}\fl
  \mu_{ev}(t) 
{}-\frac2{d_v}\sum_{e=1}^{d_v}\sint \pd{G_0}x(t,0;s,L_e)\mu_{et}(s)
+\frac2{d_v}\sum_{e=1}^{d_v}\sint \pd{^2G_0}{x\, \partial y}
 (t,0;s,L_e)\nu_{et}(s) 
 \nonumber\\
=\frac2{d_v}\pd{G_0}x(t,0;s_0,y_0) \sum_{e=1}^{d_v} \delta_{ee_0}\,.
\label{a52}\end{eqnarray}
Observe now that
 \be
 \pd{^2G_0}{x\, \partial y} (t,x;s,y) =
\frac1{2\pi}\, {(t-s)^2-(x-y)^2\over [(t-s)^2+(x-y)^2]^2 }
 = -\, \pd{^2G_0}{s\, \partial t} 
 \label{a53}\ee
 and 
 \be
 \pd{G_0}t(t,x;s,y) = -\,\frac1{2\pi}\, {t-s\over (t-s)^2+(x-y)^2}
 = -\,\pd{G_0}s\,.
 \label{a54}\ee
Some integrations by parts then put the system into the form
 \begin{eqnarray}\fl
  \nu_{ev}'(t)+2\sint \pd{G_0}t(t,0;s,L_e)\mu_{et}(s)
 \nonumber \\
\null\qquad{}-2\sint \pd{G_0}{y} (t,0;s,L_e)\nu_{et}'(s) 
+2 \pd{G_0}{t}(t,0;s_0,y_0)\delta_{ee_0}  \nonumber \\
= \mbox{same with $e\to f$}, 
\label{a55} \end{eqnarray}                               
 \begin{eqnarray}\fl
  \mu_{ev}(t) 
{}-\frac2{d_v}\sum_{e=1}^{d_v}\sint \pd{G_0}x(t,0;s,L_e)\mu_{et}(s)
+\frac2{d_v}\sum_{e=1}^{d_v}\sint \pd{G_0}{t}
 (t,0;s,L_e)\nu_{et}'(s) 
 \nonumber\\
=\frac2{d_v}\pd{G_0}x(t,0;s_0,y_0) \sum_{e=1}^{d_v} \delta_{ee_0}\,.
\label{a56}\end{eqnarray}
Recall that $e$ and $f$ are edges attached to the same vertex $v$,
  ``$t$'' refers to the other vertex of the edge in question,
  $e_0$ is the edge containing the source point $y_0\,$, and the 
  derivatives are in the direction away from~$v$.
 Because of constraints \eref{o48} and \eref{o49},
  effectively there are $\sum_v d_v$ independent components of $\nu'$ and 
$\mu$ (twice the number of edges in the graph), although 
superficially the number of unknowns is twice that, and the number 
of equations even larger if all pairs $\{e,f\}$ are counted.
 Again one can conclude by induction and Appendix A that all 
 terms in the $\nu'$ and $\mu$ series are of the same type as their 
respective source terms in \eref{a55}--\eref{a56}.

 Besides the redundancy of components,
there are two technical obstacles to showing that \eref{a55}--\eref{a56}
 manifests an integral operator of norm at most~$1$.
The first is that ``time'' derivatives of $G_0$ [\eref{a54}] do not 
fall off as fast in the ``time'' variables as space derivatives 
[\eref{deriv}] do.
 However, one can hope to apply the generalized Young inequality
 \cite[Theorem 0.10]{folland} with the weight function 
 $\rho(s) = (1+|s|)^{-1}$ in all the integrals.
 With respect to that measure, every integral kernel in 
 \eref{a55}--\eref{a56} is, with respect to the variables $t$ and 
$s$ separately, an $L^1$ function with norm at most~$1$ (in view of
 \eref{A7} and some trivial estimates).
 It follows that each term defines a mapping of norm at most~$1$ in 
the Banach space $L^p_\rho$ ($1\le p\le \infty$) of functions such that
 \be
  \int_{-\infty}^\infty |f(s)|^p \,\rho(s)^{1-p}\,ds < \infty
 \label{lp} \ee
(if $p=\infty$, the space of functions such that 
 $|f(s)|/\rho(s)$ is bounded almost everywhere).

 The second obstacle is that each equation contains two such terms, 
not one, and after redundant components are eliminated from 
 \eref{a55} via \eref{o49} the number of terms will be still 
larger.
 Therefore, the obvious bound on the one-sided $L^1$ norms of the 
 matrix-valued integral kernel is larger than~$1$,
 and one cannot conclude by this method that the series even 
marginally converges.

\section{Finite temperature} \label{app:temper}

The equilibrium state at temperature $T$ should be described by 
replacing the usual cylinder kernel by a function  periodic in $t$ 
with period $\beta=1/T$.
In general, this construction can be implemented by replacing 
$G_0$ by
\be
 G_T(t,\mathbf{x};s,\mathbf{y})=
  \sum_{N=-\infty}^\infty G_0(t,\mathbf{x};s-N/T,\mathbf{y}),
\label{A9}\ee 
 or by solving the partial differential equation with periodic
 boundary condition directly, or by summing the 
appropriate eigenfunction expansion with a Boltzmann weighting factor.

 In total dimension 2, where $G_0$ is given by \eref{o1},
the literal \eref{A9}  fails because of divergence.
 However, by making the arbitrary constants in \eref{o1} dependent 
on~$N$ one can arrive at
 \be
 G_T= -\,\frac1{4\pi} \ln\bigl[\cosh\bigl(2\pi T(x-y)\bigr)
-\cos\bigl(2\pi T(t-s)\bigr)\big] +C,
 \label{A10}\ee
whose first derivatives are
\begin{eqnarray}
 \pd {G_T}t &= -\,\frac T2\, 
{\sin \bigl(2\pi T(t-s)\bigr) \over
 \cosh\bigl(2\pi T(x-y)\bigr) - \cos\bigl(2\pi T(t-s)\bigr) }\,,
\label{A11}\\
\pd {G_T}x &= -\,\frac T2\, 
{\sinh \bigl(2\pi T(x-y)\bigr) \over
 \cosh\bigl(2\pi T(x-y)\bigr) - \cos\bigl(2\pi T(t-s)\bigr) }\,.
\label{A12} \end{eqnarray}
One can check the correctness of \eref{A10} by showing that it 
satisfies all the necessary conditions:
\begin{enumerate}
\item Periodicity in $t$ with minimal period $1/T$.
\item Delta-function initial singularity: 
  When $t-s$ and $\mathbf{x}-\mathbf{y}$  are small,
  $G_T \approx G_0\,$, which is known to have the correct behavior.
\item Asymptotic behavior at spatial infinity:
 logarithmic growth, just like $G_0\,$. 
\item  Partial differential equation 
\be
 \pd {^2G_T}{t^2} +  \pd {^2G_T}{x^2}= 0. 
 \label{A13}\ee
\end{enumerate}
 Moreover, as in  \eref{a53} there holds
  \be
 \pd{^2G_0}{x\, \partial y} 
 = -\, \pd{^2G_0}{s\, \partial t} \,.
 \label{a53b}\ee

 \textsl{Remark:\/} 
Alternatively, to deduce \eref{A12} from \eref{A9}, use the identity
 \begin{eqnarray}
  \sum_{N=-\infty}^\infty \frac 1 {a^2+ (b+N)^2} &=
\frac{\pi}{2a} \{\coth[\pi(a+ib)] + \coth[\pi(a-ib)]\} 
 \nonumber \\
&= \frac{\pi}a \, {\sinh(2\pi a)\over \cosh(2\pi a) - \cos(2\pi b)}
\,.
 \label{A14}\end{eqnarray}
(The first version of this formula was listed in \cite{fuliowa}.
 The simpler final form follows by elementary identities.)
Going backward (a Mittag-Leffler expansion),
  one can get the analogous formula \eref{A11},
  and then \eref{A10} is easy to guess.

 \end{document}